# BinarEye: An Always-On Energy-Accuracy-Scalable Binary CNN Processor With All Memory On Chip In 28nm CMOS

Bert Moons[*], Daniel Bankman[†], Lita Yang[†], Boris Murmann[†] and Marian Verhelst[*]
[*]ESAT-MICAS, KU Leuven, Leuven, Belgium
[†]Stanford University, Stanford, CA, USA

*Abstract*—This paper introduces BinarEye: the first digital processor for always-on Binary Convolutional Neural Networks. The chip maximizes data reuse through a Neuron Array exploiting local weight Flip-Flops. It stores full network models and feature maps and hence requires no off-chip bandwidth, which leads to a 230 1b-TOPS/W peak efficiency. Its 3-levels of flexibility – (a) weight reconfiguration, (b) a programmable network depth and (c) a programmable network width - allow trading energy for accuracy depending on the task's requirements. BinarEye's full-system input-to-label energy consumption ranges from 14.4uJ/f for 86%/CIFAR-10 and 98%/owner recognition down to 0.92uJ/f for 94%/face detection at up to 1700 frames per second. This is 3-12-70× more efficient than the state-of-the-art at on par accuracy.

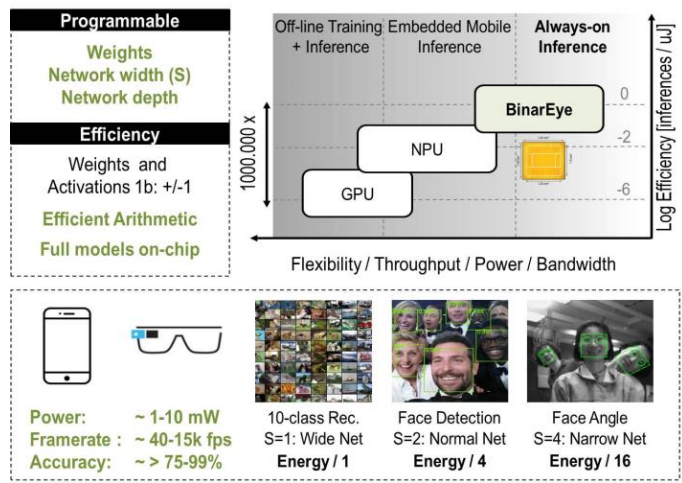

**Figure 1:** Flexible Always-On BinaryNet Inference on BinarEye

## I. INTRODUCTION

Convolutional Neural Networks (CNN) are state-of-the-art Deep Learning algorithms essential in achieving high accuracy in Computer Vision applications. Unfortunately, their performance comes at a high cost in terms of computational energy- and bandwidth, making it difficult to deploy them on battery-constrained embedded systems. Because of these challenges, CNNs have not yet been deployed on always-on mobile platforms such as smartphones, where an always-on gesture-, face-, owner- and angle-detector can be used as a wake-up sensor for its screen or application processor. Fig. 1 shows embedded GPU's and general Neural Processing Units (NPU) [2, 3, 4, 5] consume too much energy for this task, both in their computations and in their off-chip DRAM access, while classical machine learning methods [6] are either inaccurate or lack the flexibility to cover multiple tasks. This paper introduces BinarEye: a CNN processor optimized for BinaryNets [1]: Neural Networks with weights and activations constrained to +1/-1. BinarEye enables and exploits (1) extremely efficient hardware through maximal data reuse in a memory-like array, (2) storing full models and feature maps on-chip, requiring no off-chip bandwidth and (3) flexibility on three levels: reconfigurable weights, a programmable network depth and a reconfigurable network width allowing a large energy-vs-accuracy trade-off. Because of this, BinarEye can map a wide range of applications, while offering a complete Input-to-Label, full-system efficiency up to 145 1b-TOPS/W. This allows running 1k inferences/s of 125M operations each at 1mW or 1uJ/inference, while still achieving >94% precision in face detection and >90% accuracy in multiple other tasks. BinarEye outperforms [2-6] up to 70× on CIFAR-10 at a slightly lower accuracy and 10× the throughput and by 3.3-to-12× on face detection at iso-accuracy. Because of its high energy efficiency and low-power operation, BinarEye can be used as an always-on visual wake-up sensor.

## II. PROCESSOR ARCHITECTURE

Fig. 2 shows the top-level architecture and basic mode of operation of this chip. It consists of a Flip-Flop based array enabling in-memory-like compute of 64 reconfigurable binary neurons, surrounded by a total of 259kB of weight SRAM on the north- and south side storing full models and 2 × 32kB of feature SRAM on the west- and east side each storing full feature maps. BinarEye is optimized to perform F×C×k×k CNN-layers, where F (# output channels), C (# input channels) can vary and k×k is fixed to 2×2 which is sufficient for all tested tasks. Every CNN-layer starts by preloading (LD) all 64 neurons' weights from the north and south weight SRAM into a set of local flip-flops in the Neuron Array. This is done such that the weights can be loaded once from SRAM and reused throughout a convolution. Once preloaded, all neurons perform a parallel stride-1 convolution (CONV) on the same input activations. By enforcing this structural regularity, the physical architecture of this processor maximally exploits the locality of the CNN algorithm. In each convolution step on a 2×2 patch, memory bandwidth to the activation SRAM is minimized by fetching only 2 feature bits per step and storing them locally at the edge of the Neuron Array. This is possible as 2 features can always be reused from the previous step. Each of the 64 neurons then outputs one activation, which is stored in the east (west) activation memory. As all weights and activations are constrained to +1/-1, the neuron dot product ($A_{in} \cdot W + b = a_{out}$) is performed as a number of parallel XNOR-operations, followed by a bit-count and a binary comparator [1]. The chip also contains logic and 5kB of SRAM for binary fully-connected (FC) layers, enabling it to output classification labels without the need of any external blocks. In general, storing all model weights in on-chip SRAM offers a first level of flexibility in BinarEye, as the weights can be retrained for different applications.

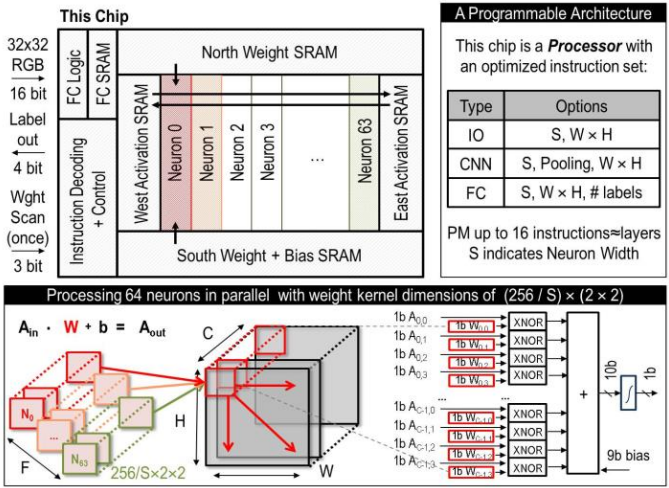

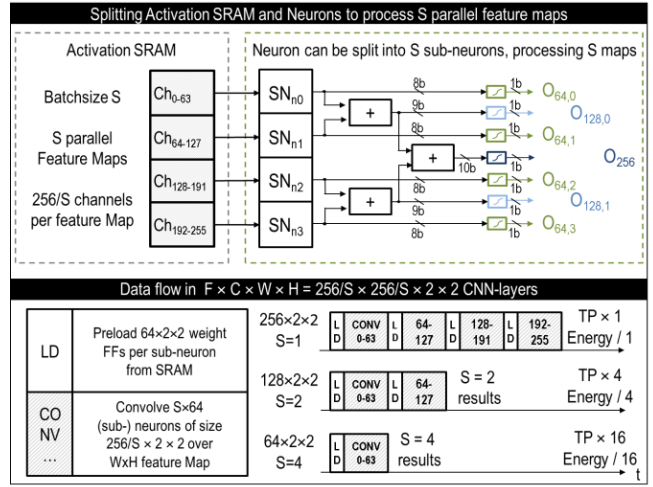

**Figure 2:** High-level Chip-overview

A second level of flexibility in BinarEye is its ability to program the network depth. This is enabled through a programmable controller which decodes custom instructions for input-output layers (IO), CNN-layers (CNN) and FC-layers (FC), through a program memory that can store up to 16 of such instructions. Fig. 2 shows BinarEye's instructions support streamed Max-Pooling in CNN-layers, a configurable W×H up to 32×32 and up to 10 classes in its FC-layer.

A third level of flexibility in BinarEye is its ability to modulate the network width, even in a rigid Flip-Flop based memory-like array. Fig. 3 shows how the width of the CNN-layers (F) and the number of channels of the input map (C) can be modulated with the variable degree of parallelism or batch size S as F=C=256/S. Varying S hence allows trading energy in exchange for modeling capacity or accuracy of the network. This is implemented by splitting neurons into 4 sub-neurons, each processing a 64×2×2 dot-product on 64 channels of the activation SRAM. Depending on S, the intermediate outputs of the sub-neurons are combined into S output features. BinarEye can hence either perform F×C×k×k

**Figure 3:** Programming Neuron width to trade-off energy for network accuracy.

= 256×256×2×2 (S=1), 128×128×2×2 (S=2) or 64×64×2×2 (S=4) layers on one 256×W×H (S=1), two 128×W×H (S=2) or four 64×W×H (S=4) feature maps in parallel. This has a significant impact on throughput and classification energy. If S=1, a layer contains 256 neurons (F=C=256), which means 4 LD-CONV phases are required to process all neurons. If S=4, a layer contains 64 (F=C=64) neurons and requires only 1 LD-CONV phase to process 4 input maps. Here, throughput and energy are improved quadratically with $S^2$=16, at the cost of reduced modeling capacity or classification accuracy. This trade-off is further quantified in Fig. 5.

The Flip-Flop based memory-like Neuron Array discussed here is identical to an in-memory compute architecture as in [7, 8], where data-movement is minimized by performing computations as close to nonvolatile [7] or SRAM [8] bit-cells as possible. All discussed architectural techniques used to turn this array into a flexible Input-to-Label programmable CNN processor can hence also be applied to such in-memory compute macros. The proposed surrounding circuitry used to control and update the neuron/memory array is crucial to any in-memory solution. In BinarEye however, all weights are locally stored in Flip-Flops rather than S- or nonvolatile RAM. As such, it requires no sense-amplifiers, data-independent bit-line pre-charging or charge-pumped writes and hence minimizes energy consumption if the registers are clock gated. Additionally, by implementing the array using standard-cells, the design can be done in a digital tool-flow requiring no custom design intervention. It can hence be reused across CMOS technology generations and requires no calibration. RAM-based in-memory solutions [7, 8] and the less flexible mixed-signal Binary CNN processor [9] built on the BinarEye architecture do require custom design and calibration.

III. PROCESSOR PERFORMANCE AND APPLICATIONS

*A. Benchmark Network Performance*

To demonstrate the chip's flexibility and the effectiveness of its different modes, Fig. 4 shows measurement results for the layers of a typical 9-layer benchmark network for always-on imaging. The first layer receives a 32×32 7-bit RGB input and transforms it into a 256×31×31 binary feature map in 500M binary operations at core-efficiencies up to 230TOPS/W at 6MHz and 352GOPS. Throughout the network, the core-efficiency drops with smaller W×H feature maps, as the relative LD-time increases compared to the CONV-time. Although BinarEye is not optimized for FC-layers, it does achieve sota efficiencies up to 1.5TOPS/W. All measurements are at room temperature at optimal supply voltages from 0.66-to-0.9V and frequencies ranging from 1.5-48MHz.

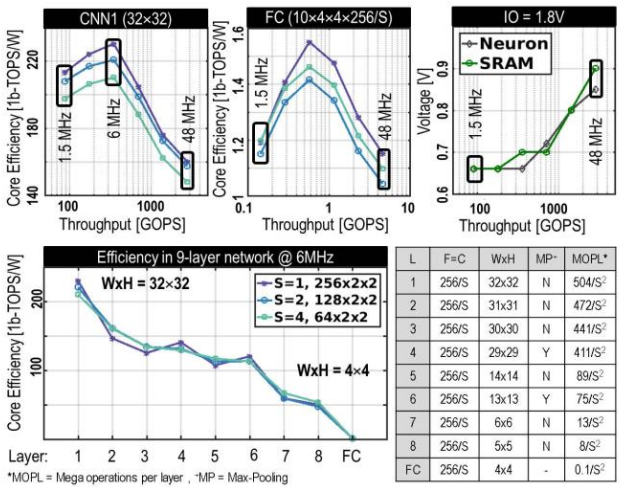

**Figure 4:** Measured core performance of separate instructions in BinarEye

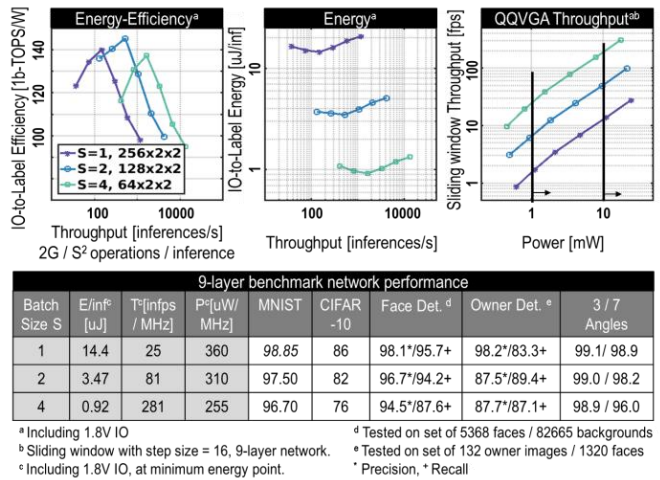

**Figure 5:** IO-to-Label performance of BinarEye.

### B. Application level performance

The system level, IO-to-Label (I2L) performance and wide applicability of BinarEye are illustrated in Fig. 5. Here, the accuracy of several benchmarks is given as a function of S together with the energy-consumption, throughput and power for the 9-layer network of Fig. 4. All data sets and training scripts are available online at https://github.com/BertMoons/ [10]. BinarEye can efficiently process several wake-up tasks relative in the context of wearable devices, among others. When S=1 at 150 inferences/s and 14.4 uJ/inference, the chip achieves up to 86% on CIFAR-10, 98.1% accuracy / 95.7% recall on Face- and 98.2% accuracy / 83.3% recall on Owner Detection. If a slightly lower accuracy is allowed, the chip can scale up to 1700 inferences/s at its minimum energy point, consuming 0.92 uJ/inference at 94.5% precision on Face Detection. Once a face or owner is detected, the chip can reliably recognize up to 3-7 angles of the face, which can allow a mobile device to accurately track the relative position of a user's face relative to its own screen. If BinarEye is used as a detector in a sliding window fashion on QQVGA images, it achieves 1-20 fps at 1mW and 15-200fps at 10mW using a stride of 16 pixels. At 1mW, the chip hence offers up to 33-days of always operation on a typical 810mWh AAA battery.

### C. Physical implementation

Fig. 6 shows BinarEye's implementation in a 28nm CMOS technology. This implementation requires 1.4mm² of active area to lay-out the Neuron Array, the north- and south-side Weight SRAMs, the west- and east-side Activation SRAMs and all control and Fully-connected layer logic. As a full-die solution, BinarEye requires only 2mm² and two separate supply voltages, one for all memories and control logic and one for the Neuron Array.

### IV. COMPARISON WITH THE STATE-OF-THE-ART

Table 1 is a comparison with recent CNN and BinaryNet chips. Due to (1) its efficient arithmetic, (2) complete on-chip network storage and (3) its three-level flexibility, BinarEye outperforms [2-6] in energy per inference. Fully binary [3] is outperformed up to 2× in peak throughput and up to 30× in energy-efficiency at 1.4TOPS, as it does not exploit data reuse. For MNIST, a narrow 5-layer network running on BinarEye is 7.3% more accurate than [3], while using 1.33× less energy. On CIFAR-10, it consumes 14.5uJ/inference at 86% accuracy. This is a 70× advantage over [2] if IO power is included, at a slightly lower accuracy but at 10× higher throughput. Compared to a 1-chip version of IBM TrueNorth [5], BinarEye is 11.4× more energy-efficient at 3% higher accuracy and a 13× lower EDP. In Face Detection, BinarEye consumes 12× less than fixed-function [6] and 3.3× less than [4] at iso-accuracy. Furthermore, BinarEye achieves this at extremely low power: at Emin, it consumes only 1.6-2.2 mW, a 4-100× advantage over [2, 3, 4, 5]. This shows BinarEye can be used for always-on visual recognition with a scarce power budget, in applications ranging from face- and owner detection to 10-class object recognition.

### V. CONCLUSION

This paper introduces BinarEye: the first fully digital processor for always-on Binary CNNs. It achieves always-on operation on a number of tasks due to three key innovations. (1) The chip is a flexible memory-like binary Neuron Array with local weight flip-flops maximizing weight reuse in CNNs. (2) It requires no off-chip bandwidth due to its large on-chip memory, which enables storing full models and feature maps. (3) It is flexible on three levels: reprogrammable weights, a custom instruction set to program network depth and a reconfigurable network width allowing trading energy for accuracy depending on the specific requirements of a task. Because of this, it achieves up to 230 1b-TOPS/W peak efficiency and up to 145 1b-TOPS/W Input-to-Label (I2L) efficiency with all overheads included. Its full-system input-to-label energy consumption ranges from 14.4uJ/f for 86%/CIFAR-10 and 98%/owner recognition down to 0.92uJ/f for 94%/face detection at up to 1.7 kfps. This is 3-12-70× more efficient than the state-of-the-art at on par accuracy, hereby enabling always-on recognition tasks on battery constrained wearable platforms.

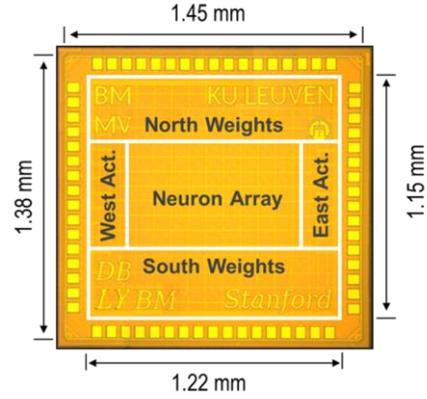

**Figure 6:** Die Micrograph of BinarEye.

**Table 1:** Comparison of Low-Precision CNN Processors on multiple benchmarks. *Core efficiency is without IO-power. † is with IO-power included. ‡ Energy on Conv-layers only. α Numbers for the 9-layer benchmark of Fig. 4.

|  | [2] TCAD '17 | [3] VLSI '17 | [4] ISSCC'17 | [5] IBM TrueNorth | [6] ISSCC'17 | **This work** |
|---|---|---|---|---|---|---|
| Technology | 65nm CMOS | 65nm CMOS | 28nm FDSOI | 28nm CMOS | 65nm CMOS | **28nm CMOS** |
| Frequency [MHz] | 27 - 480 | 100 - 400 | 25-200 | - | 25 | **1.5 - 48** |
| Supply [V] | 0.6 - 1.2 | 0.55 - 1 | 0.6-1 | 1 | 2.5 | **0.66 – 0.9** |
| Active Area [mm²] | 2.2 | 3.9 | 1.87 | 430 | - | **1.4** |
| # of MACs | - | 1728 | 256 - 1024 | - | - | **65536** |
| Gate Count [NAND-2] | 1.33M | - | 1.95M | 5.4B transistors | - | **1.3M** |
| On-Chip Memory | 9.2kB SCM | 51kB SRAM | 148kB SRAM | 51.2 MB SRAM | 80×20 An. | **328kB SRAM** |
| #layers, #filters, sizes [-] | All, All, <7×7 | 13, -, - | All, All, All | All, All, All | - | **1-16, 64-256, 2×2** |
| Supported Networks [-] | CNN | DNN | CNN | CNN | Haar-Filters | **CNN + DNN** |
| Precision [bits] | 1b × 12b | 1-1.6b | 1-16b | 1-1.6b | Analog | **1b** |
| Performance [GOPS] | 15 – 377 | 345 – 1380 | 12 – 408 | - | - | **90 – 2800** |
| Core* Eff. [TOPS/W] | 58.6 - 9.6 | 6 – 2.3 | 10 – 0.3 | - | - | **230 - 145** |
| I2L† Eff. [TOPS/W] | 0.98 - 0.87 | - | - | - | - | **145 – 95** |

| Acc. [%] | $E_{min}$[uJ/f] | [%] | ‡* | †† | [%] | * | [%] | ‡* | [%] | * | [%] | * | [%] | Core* | I2L† | S |
|---|---|---|---|---|---|---|---|---|---|---|---|---|---|---|---|---|
| MNIST |  | - | - | - | 90.1 | 0.28 | - | - | - | - | - | - | 97.4 | 0.2 | 0.21 | 4 |
| CIFAR-10 |  | 91.7 | 21 | 1k | - | - | - | - | 83.4 | 164 | - | - | 86 | 13.82 | 14.4 | 1 |
| Face Detection |  | - | - | - | - | - | 94 | 3 | - | - | >95 | 11.8 | 94.5 | 0.89 | 0.92 | 4 |
| Owner Detection |  | - | - | - | - | - | - | - | - | - | - | - | 98.2 | 13.82 | 14.4 | 1 |
| 7 Face Angles |  | - | - | - | - | - | - | - | - | - | - | - | 98.2 | 3.4 | 3.47 | 2 |
| Mode S |  | - |  |  | - |  | - |  | - |  | - |  | 1α | 2α | 4α |  |
| Operations / Net [-] |  | 1.2G |  |  | 1.3M |  | 12.4M |  | - |  | - |  | 2G | 0.5G | 0.12G |  |
| Inf/s @ Emin, Net [-] |  | 15.6 |  |  | 205k |  | 2.2k |  | 1249 |  | 1 |  | 0.12k | 0.5k | 1.7k |  |
| EDP @Emin, Net[uJs] |  | 64.1†† |  |  | 1.4e-6* |  | 1.4e-3‡* |  | 0.131 |  | - |  | 1e-2† | 7e-3† | 5e-4† |  |
| P @ Emin, Net [mW] |  | 15.760†† |  |  | 50* |  | 6.4‡* |  | 204.4 |  | - |  | 2.2† | 1.8† | 1.6† |  |